# "Illusion of Control" in Minority and Parrondo Games

J.B. Satinover[1,a] and D. Sornette[2,b]

[1]Laboratoire de Physique de la Matière Condensée, CNRS UMR6622 and Université des Sciences, Parc Valrose, 06108 Nice Cedex 2, France

[2]Department of Management, Technology and Economics, ETH Zurich, CH-8032 Zurich, Switzerland

**Abstract:** Human beings like to believe they are in control of their destiny. This ubiquitous trait seems to increase motivation and persistence, and is probably evolutionarily adaptive [1,2]. But how good really is our ability to control? How successful is our track record in these areas? There is little understanding of when and under what circumstances we may over-estimate [3] or even lose our ability to control and optimize outcomes, especially when they are the result of aggregations of individual optimization processes. Here, we demonstrate analytically using the theory of Markov Chains and by numerical simulations in two classes of games, the Minority game [4,5] and the Parrondo Games [6,7], that agents who optimize their strategy based on past information actually perform worse than non-optimizing agents. In other words, low-entropy (more informative) strategies under-perform high-entropy (or random) strategies. This provides a precise definition of the "illusion of control" in set-ups a priori defined to emphasize the importance of optimization.

## 1 Introduction

The success of science and technology, with the development of ever more sophisticated computerized integrated sensors in the biological, environmental and social sciences, illustrate the quest for control as a universal endeavor. The exercise of governmental authority, the managing of the economy, the regulation of financial markets, the management of corporations, and the attempt to master natural resources, control natural forces and influence environmental factors all arise from this quest. Langer's phrase, "illusion of control" [3] describes the fact that individuals appear hard-wired to over-attribute success to skill, and to underestimate the role of chance, when both are in fact present. Whether control is genuine or merely perceived is a prevalent question in psychological theories. The sequel presents two rigorously controlled mathematical set-ups demonstrating generic circumstances in which optimizing agents perform worse than non-optimizing or random agents.

## 2 Minority Games

We first study Minority games (MGs), which constitute a sub-class of market-entry games. MGs exemplify situations in which the "rational expectations" mechanism of standard economic theory fails. This mechanism in effect asks, "what expectational model would lead to collective actions that would on average validate the model, assuming everyone adopted it,"?[8]. In minority games, a large number of interacting decision-making agents, each aiming for personal gain in an artificial universe with scarce resources, try to anticipate the actions of others on the basis of incomplete information. Those who subsequently find themselves in the minority group gain. Therefore, expectations that are held in common negate themselves, leading to anti-persistent behavior both for the aggregate behavior and for individuals. Minority games have been much studied as repeated games with expectation indeterminacy, multiple equilibria and inductive optimization behavior.

[a] jsatinov@princeton.edu
[b] dsornette@ethz.ch

Consider the Time-Horizon MG (THMG), where N players have to choose one out of two alternatives at each time step based on information represented as a binary time series A(t). Those who happen to be in the minority win. Each agent is endowed with S strategies. Each strategy gives a prediction for the next outcome A(t) based on the history of the last m realizations A(t-1), ..., A(t-m) (m is called the memory size of the agents). Each agent holds the same number S of (in general different) strategies among the total number $2^{2^m}$ of strategies. The S strategies of each agent are chosen at random but once and for all at the beginning of the game. At each time t, in the absence of better information, in order to decide between the two alternatives for A(t), each agent uses her most successful strategy in terms of payoff accumulated in a rolling window of finite length $\tau$ up to the last information available at the present time t. This is the key optimization step. If her best strategy predicts A(t)=+1 (resp. -1), she will take the action $a_i(t)$ = - 1 (resp. +1). The aggregate behavior A(t) = $\sum_{i=1}^{N}$ $a_i(t)$ is then added to the information set available for the next iteration at time t+1. The corresponding instantaneous payoff of agent i is given by – sign[$a_i(t)$ A(t)] (and similarly for each strategy for which it is added to the $\tau$-1 previous payoffs). As the name of the game indicates, if a strategy is in the minority ($a_i(t)$ A(t) < 0), it is rewarded. In other words, agents in MG try to be anti-imitative. The richness and complexity of minority games stem from the fact that agents strive to be different. Previous investigations have shown the existence of a phase transition between an inefficient regime and an uncorrelated phase as the parameter $2^m/N$ is increased, with the size of the fluctuations of A(t) (as measured by its normalized variance $\sigma^2/N$) falling below the random coin-toss limit for large *m*'s when agents always use their highest scoring strategy [5]. The phenomenon discussed here is not directly related to these effects as it is observed in all regimes.

Our main result may be stated concisely from the perspective of utility theory: Throughout the space of parameters (N, m, S, $\tau$), the mean payoff of strategies not only surpasses the mean payoff of supposedly-optimizing agents, but the respective cumulative distribution functions (CDF) of payoffs show a first-order stochastic dominance of strategies over agents. Thus, were the option available to them, agents would behave in a risk-averse fashion (concave utility function) by switching randomly between strategies rather than optimizing. This result generalizes when comparing optimizing agents with S>1 strategies with agents having only one strategy (or equivalently m identical strategies), when the single strategies are actually implemented (which takes into account the impact of the strategy itself). The same result is also found when comparing optimizing agents with agents flipping randomly between their S strategies. Agents are supposed to enhance their performance by choosing adaptively between their available strategies. In fact, the opposite is true: The optimization method is found to be strictly a method for worsening performance!

This "illusion of control" effect is observed for all N, m, S and $\tau$. Indeed, extensive numerical simulations show that all the phenomena we derive for the THMG with the simplest parameter values are found in the MG with arbitrary parameters. We use the Markov chain formalism for the THMG [8,9] to obtain the following theoretical prediction for the average gains $\Delta W_{Agent}$ and $\Delta W_{Strategy}$, respectively of optimizing agents and of their strategies [10]:

$$\langle \Delta W_{Agent} \rangle = \tfrac{1}{N} |\vec{A}_D| \cdot \vec{\mu} \qquad (1)$$

and

$$\langle \Delta W_{Strategy} \rangle = \tfrac{1}{2N}\left(\hat{\mathbf{s}}_\mu \cdot \vec{\kappa}\right) \cdot \vec{\mu} \qquad (2)$$

where the brackets denote a time average. $\mu$ is a $(m+\tau)$-bit "path history" (sequence of 1-bit states); $\vec{\mu}$ is the normalized steady-state probability vector for the history-dependent $(m+\tau) \times (m+\tau)$ transition matrix $\hat{\mathbf{T}}$, where a given element $T_{\mu_t, \mu_{t-1}}$ represents the transition probability that $\mu_{t-1}$ will be followed by $\mu_t$; $\vec{A}_D$ is a $2^{(m+\tau)}$-element vector listing the particular sum of decided values of A(t) associated with each path-history; $\hat{\mathbf{s}}_\mu$ is the table of points accumulated by each strategy for each path-history; $\vec{\kappa}$ is a $2^{(m+\tau)}$-element vector listing the total number of times each strategy is represented in the collection of N agents. As shown in the supplementary material, $\hat{\mathbf{T}}$ may be derived from $\vec{A}_D$, $\hat{\mathbf{s}}_\mu$ and $\vec{N}_U$, the number of undecided agents associated with each path history. Thus agents' mean gain is determined by the non-stochastic contribution to A(t) weighted by the probability of the possible path histories. This is because the stochastic contribution for each path history is binomially distributed about the determined contribution. Strategies' mean gain is determined by the change in points associated with each strategy over each path-history weighted by the probability of that path.

We find excellent agreement between the numerical simulations and the analytical predictions (1) and (2) for the THMG. For instance, for m=2, S=2, $\tau$=1 and N=31, $\langle \Delta W_{Agent} \rangle$=-0.22 for both analytic and numerical methods (payoff per time step averaged over time and over the optimizing agents) compared with $\langle \Delta W_{Strategy} \rangle$=-0.06 also for both analytic and numerical methods (corresponding payoff for individual strategies). In this numerical example, the average payoff of individual strategies is larger than for optimizing agents by 0.16 units per time step. The numerical values of the predictions (1) and (2) are obtained by implementing each agent individually as a coded object.

Taking into account the impact of strategies modifies these results quantitatively but not qualitatively. The mean per-agent per-step payoff $\langle \Delta W_{Non-Opt} \rangle$ accrued by non-optimizing agents (because they have only one fixed strategy, or equivalently their S strategies are identical) is larger than the payoff $\langle \Delta W_{Agent} \rangle$ of optimizing agents. In general, this comparative advantage decreases with their proportion. For example, with m=2, S=2, $\tau$=1 and N=31, and 2500 random initializations and n optimizing agents, $\langle \Delta W_{Non-Opt} \rangle - \langle \Delta W_{Agent} \rangle$= (2.380, 2.270, 2.289, 2.275, 2.145, 2.060, 2.039, 1.994, 1.836, 1.964) $\times 10^{-3}$ for n=1, 2, …, 10. More generally, the following ordering holds: payoff of individual strategies > payoff of non-optimizing agents > payoff of optimizing agents. The first inequality is due to the fact that not all individual strategies are implemented and the theoretical payoff of the non-implemented strategies does not take into account what their impact would have been (had they been implemented). Implementation of a strategy tends to decrease its performance (this is similar to the market impact of trading strategies in financial markets associated with slippage and market friction). Non-optimizing agents by definition always implement their strategies. However, the higher payoff of non-optimizing compared with optimizing agents shows that the illusion-of-control effect is not due to the impact mechanism, but is a genuine observable effect.

The amplitude of the illusion-of-control effect highlights the distinction between optimizing agents with S maximally distinct strategies (in the sense of Hamming distance) and non-optimizing agents with S identical strategies. It is helpful to generalize this dichotomy by characterizing the degree of similarity between the S strategies of a given agent using the Hamming distance $d_H$ between strategies (the Hamming distance between two strings of equal length is the number of positions for which the corresponding symbols are different). Non-optimizing agents with S identical strategies corresponds to $d_H=0$. In contrast, optimizing agents with S maximally distinct strategies have large $d_H$'s. Since agents with $d_H=0$ out-perform agents with large $d_H$, it is natural to ask whether the ranking of $d_H$ could be predictive of the ordering of agents' payoffs. A non-zero Hamming $d_H$ implies that there are at least two strategies among the S strategies of the agent which are different. But, if $d_H$ is small, the small difference between the S strategies makes the optimization only faintly relevant and one can expect to observe a payoff similar to that of non-optimizing agents, therefore larger than for optimizing agents with large $d_H$'s. This intuition is indeed confirmed by our calculations: the average payoff per time step is a decreasing function of $d_H$, as originally discussed in [11,12].

The illusion-of-control effect suggests that the initial set-up of MG in terms of S fixed strategies per agent is evolutionarily unstable. It is thus important to ask what happens when agents are allowed to replace strategies over time based on performance. A number of authors have investigated this issue, adding a variety of longer-term learning mechanisms on top of the short-term adaptation that constitutes the basic MG [12-18]. Inter alia, Ref.[13] demonstrates that if agents are allowed to replace strategies over time based on performance, they do so by ridding themselves of those composed of the more widely Hamming-distant tuples. Agents that start out composed of identical strategies do not change at all; those composed of strategies close in Hamming space change little. Similarly, the authors of [12] explicitly fixed agents with tuples of identical strategies and found they performed best. Another important finding in [12] is that the best performance attainable is equivalent to that obtained by agents choosing their strategies at random. Note that learning only confers a *relative* advantage. In general, agents that learn out-perform agents that don't. This is certainly true for this privileged subset of agents among standard ones. But the performance of learning agents approaches a maximum most closely attained by agents where the hamming distance between strategies is 0. These agents neither adapt (optimize) nor learn. One might say that when learning is introduced, the system learns to rid itself of the illusory optimization method that has been hampering it.

There are exceptions, of course. Carefully designed privileges and certain kinds of learning can yield superior results for a subset of agents, and occasionally for all agents. But the routine outcome is that both optimization and straightforward learning cannot improve on simple chance. The fact that the optimization method employed in the MG yields the opposite of the intended consequence, and that learning eliminates the method, leads to an important question. We pose it carefully so as to avoid introducing either privileged agents or learning: Is the illusion-of-control so powerful in this instance that inverting the optimization rule could yield equally unanticipated and opposite results? The answer is yes: If the fundamental optimization rule of the MG is symmetrically inverted for a limited subset of agents who choose their worst-performing strategy instead of their best, those agents systematically outperform both their strategies and other agents. They also can attain positive gain. Thus, the intuitively self-evident control over outcome proffered by the MG "optimization" strategy is most strikingly shown to be an illusion. Even learning and evolutionary strategies generally at best rid the system of any optimization method altogether. They do not attain the kind of results obtained simply by allowing some agents to reverse the method altogether. We discuss elsewhere the phenomena that arise as the proportion of agents choosing their best performing strategy and of agents choosing their worst performing strategy are varied for different parameters of the MG. We emphasize here only the fact that extensive numerical studies confirm that the phenomenon here indicated persist over a very wide range of parameters for both the MG and the THMG. Hence, having a portfolio of S strategies to choose from is counter-productive: (diversification + optimization) performs on average worse than a single fixed strategy.

Intuitively, the illusion-of-control effect in MG results from the fact that a strategy that has performed well in the past becomes crowded out in the future due the minority mechanism: performing well in the recent past, there is a larger probability for a strategy to be chosen by an increasing number of agents, which inevitably leads to its demise. This argument in fact also applies also to all the strategies which belong to the same reduced set; their number is $2^{2^m}/2^m$, equal to the ratio of the cardinality of the set of all strategies to the cardinality of the set of reduced strategies. Thus, the crowding mechanism operates from the fact that a significant number of agents have at least one strategy in the same reduced subset among the $2^m$ reduced strategy subsets. Optimizing agents tend on average to adapt to the past but not the present. They choose an action a(t) which is on average out-of-phase with the collective action A(t). In contrast, non-optimizing agents average over all the regimes for which their strategy may be good and bad, and do not face the crowding-out effect. The crowding-out effect also explains simply why anti-optimizing agents over-perform: choosing their worst strategy ensures that it will be the least used by other agents in the next time step, which implies that they will be in the minority. The crowding mechanism also predicts that the smaller the parameter $2^m/N$, the larger the illusion-of-control effect. Indeed, for large values of $2^m/N$, it becomes more and more probable that agents have their strategies in different reduced strategy classes, so that a strategy which is best for an agent tells nothing about the strategies used by the other agents, and the crowding out mechanism does not operate. Thus, regions of successful optimization, if they occur at all, are more likely at higher values of $2^m/N$. (See supplementary material for further details.)

This leads to the conclusion that there is a profound clash between optimization on the one hand and minority payoff on the other hand: an agent who optimizes identifies her best strategy, but in so doing by her "introspection", she somehow knows the fate of the other agents, that it is probable that the other agents are also going to choose similar strategies, … which leads to their underperformance since most of them will then be in the majority. It follows then that an optimizing agent playing a standard minority game should optimize at a second order of recursion in order to win: Her best strategy allows her to identify the class of best strategies of others, which she thus must avoid absolutely to be in the minority and to win (given that other players are just optimizing at the first order as in the standard MG). Generalization to ever more complex optimizing set-ups, in which agents are aware of prior-level effects up to some finite recursive level, can in principle be iterated ad infinitum.

Actually, the game theory literature on first-entry games shows that the resulting equilibria depend on how agents learn [19]: with reinforcement learning, pure equilibria involve considerable coordination on asymmetric outcomes where some agents enter and some stay out; learning with stochastic fictitious plays leads to symmetric equilibria in which agents randomize over the entry decisions. There may even exist asymmetric mixed equilibria, where some agents adopt pure strategies while others play mixed strategies. We consider the situation where agents use a boundless recursion scheme to learn and optimize their strategy so that the equilibrium corresponds to the fully symmetric mixed strategies where agents randomize their choice at each time step with unbiased coin tosses. Consider a MG game with N agents total, $N_R$ of which employ such a fully random symmetric choice. The remaining $N_S = N-N_R$ "special" agents (with $N_R \gg N_S$) will all be one of three possible types: agents with S fixed strategies that choose their best (respectively worst) performing strategy to make the decision at the next step (referred to above as anti-optimizing) and agents with a single fixed strategy. Our simulations confirm that these three types of agents indeed under-perform on average the optimal fully symmetric purely random mixed strategies of the $N_R$ agents (see Fig. 4 of the Supplementary materials). Here, pure random strategies are obtained as optimal, given the fully rational fully informed nature of the competing agents. The particular results are sensitive to which strategies are available to the special agents and to their proportion. Their underperformance in general requires averaging over all possible strategies and S-tuples of strategies. (In the supplementary material we show sample numerical results for $N_S = 1$).

## 3 Parrondo Games

We now turn to the illusion-of-control effect in the Parrondo games (PG). The basic Parrondo effect (PE) was first identified as a game-theoretic equivalent to the directional drift of Brownian particles in a time-varying "ratchet"-shaped potential [20,21], wherein two or more individually fair (or losing) games yield a net winning outcome if alternated periodically or randomly. Consider $N > 1$ $s$-state Markov games $G_i$, $i \in \{1,2,...,N\}$, and their $N$ $s \times s$ transition matrices, $\hat{\mathbf{M}}^{(i)}$. For every $\hat{\mathbf{M}}^{(i)}$, denote its vector of $s$ winning probabilities conditional on each of the $s$ states as $\vec{\mathbf{p}}^{(i)} = \{p_1^{(i)}, p_2^{(i)}, ... p_s^{(i)}\}$ and its steady-state equilibrium distribution vector as $\vec{\Pi}^{(i)} = \{\pi_1^{(i)}, \pi_2^{(i)}, ..., \pi_s^{(i)}\}$. For each game, the steady-state probability of winning is therefore $P_{win}^{(i)} = \vec{\mathbf{p}}^{(i)} \cdot \vec{\Pi}^{(i)}$. Consider also a sequence of randomly alternating $G_i$ with individual time-averaged proportion of play $\gamma_i \in [0,1]$, $\sum_{i=1}^{N} \gamma_i = 1$. The transition matrix for the combined sequence of games is the convex linear combination $\hat{\mathbf{M}}^{(\gamma_1, \gamma_2, ..., \gamma_N)} \equiv \sum_{i=1}^{N} \gamma_i \hat{\mathbf{M}}^{(i)}$ with conditional winning probability vector $\vec{\mathbf{p}}^{(\gamma_1, \gamma_2, ..., \gamma_n)} = \sum_{i=1}^{n} \gamma_i \vec{\mathbf{p}}^{(i)}$ and steady-state probability vector $\vec{\Pi}^{(\gamma_1, \gamma_2, ..., \gamma_n)}$ (which is a complex nonlinear mixture of the $\vec{\Pi}^{(i)}$'s). The steady-state probability of winning for the combined game is therefore

$$P_{win}^{(\gamma_1, \gamma_2, ..., \gamma_N)} = \vec{\mathbf{p}}^{(\gamma_1, \gamma_2, ..., \gamma_N)} \cdot \vec{\Pi}^{(\gamma_1, \gamma_2, ..., \gamma_N)} \qquad (3)$$

A PE occurs whenever (and in general it is the case that)

$$\sum_{i=1}^{N} \gamma_i P_{win}^{(i)} \neq P_{win}^{(\gamma_1, \gamma_2, ..., \gamma_N)} \quad \text{i.e.} \quad \sum_{i=1}^{N} \gamma_i \vec{\mathbf{p}}^{(i)} \cdot \vec{\Pi}^{(i)} \neq \vec{\mathbf{p}}^{(\gamma_1, \gamma_2, ..., \gamma_N)} \cdot \vec{\Pi}^{(\gamma_1, \gamma_2, ..., \gamma_N)} \qquad (4)$$

hence the PE, or "paradox", when the left hand sides of (4) are less than zero and the right-hand sides greater.

Many variants have been studied, including capital-dependent multi-player PG (MPPG) [22,23]: At (every) time-step $t$, a constant-size subset of all participants is randomly re-selected actually to play. All participants keep individual track of their own capital but do not alternate games independently based on it. Instead this data is used to select which game the participants must use at $t$. The chosen game is the one which, given the individual values of the capital at $t-1$ and the known matrices of the two games and their linear convex combination, has the most positive expected *aggregate* gain in capital, summed over all participants. This rule may be thought of as a static optimization procedure—static in the sense that the "optimal" choice appears to be known in advance. It appears exactly quantifiable because of access to each player's individual history. If the game is chosen at random, the change in wealth averaged over all participants is significantly positive. But when the "optimization" rule is employed, the gain becomes a loss significantly greater than that of either game alone. The intended "optimization" scheme actually reverses the positive (collective) PE. The reversal arises in this way: the "optimization" rule causes the system to spend much more time playing one of the games, and individually, any one game is losing.

Here, we present a more natural illustration of the illusion-of-control: while the MG is intrinsically collective, PGs are not. Neither the capital- nor the history-dependent variations require a collective setting for the PE to appear as shown from (4). Thus, the effect is most clearly demonstrated in a single-player implementation with two games under the most natural kind of optimization rule: at time $t$, the player plays whichever game has accumulated the most points (wealth) over a sliding window of $\tau$ prior time-steps from $t-1$ to $t-\tau$. Under this rule, a "current reversal" (reversal of a positive PE) appears. By construction, the individual games $\hat{\mathbf{M}}^{(1)}$ and $\hat{\mathbf{M}}^{(2)}$ played individually are both losing; random alternation between them is winning (the PE effect (4)), but unexpectedly, choosing the previously best-performing game yields losses slightly less than either $\hat{\mathbf{M}}^{(1)}$ and $\hat{\mathbf{M}}^{(2)}$ individually: the PE is almost entirely eliminated. Furthermore, if instead the previously *worst* performing game is chosen, the player does better than either game and even much better than the PE from random game choice. Under the choose-best optimization rule, two matrices $\hat{\mathbf{M}}^{(1)}$ and $\hat{\mathbf{M}}^{(2)}$ do not form a linear convex sum. Instead, the combined game is represented by an $(s+\tau) \times (s+\tau)$ transition matrix $\hat{\mathbf{Q}}^{(1,2)}$ with conditional winning probabilities $q_j = \frac{1}{2}\left\{ p^{(1)}_{\alpha(j)}\left[1 + p^{(1)}_{\beta(j)} - p^{(2)}_{\beta(j)}\right] + p^{(2)}_{\alpha(j)}\left[1 - p^{(1)}_{\beta(j)} + p^{(2)}_{\beta(j)}\right]\right\}$ with $j = 1, 2, \ldots, 2s$ and indices $\alpha(j) = Mod[j-1, 4] + 1$, $\beta[j] = \frac{1}{2}(j - Mod[j-1, 2] + 1)$. (Under the choose-worst rule $q_j = \frac{1}{2}\left\{ p^{(1)}_{\alpha(j)}\left[1 - p^{(1)}_{\beta(j)} + p^{(2)}_{\beta(j)}\right] + p^{(2)}_{\alpha(j)}\left[1 + p^{(1)}_{\beta(j)} - p^{(2)}_{\beta(j)}\right]\right\}$). Using matrices with the same values as studied in [24,25], the one-player two-game history-dependent PE is as follows: $\hat{\mathbf{M}}^{(1)}$ and $\hat{\mathbf{M}}^{(2)}$ have respective winning probabilities $P^1_{win} = 0.494$ and $P^2_{win} = 0.495$. Alternated at random in equal proportion $(\gamma_1 = \gamma_2 = 0.5)$, $P^{\gamma_1=0.5,\gamma_2=0.5}_{win} = 0.501$. If the previously winning game is selected, $P^{best(1,2)}_{win} = 0.496$, while if the previously losing one is, $P^{best(1,2)}_{win} = 0.507$. The mechanism for this illusion-of-control effect characterized by the reversing of the PE under optimization is not the same as for the MG, as there is no collective effect and thus no-crowding out of strategies or games. As seen from (4), the PE results from a distortion of the steady-state equilibrium distributions $\vec{\Pi}^{(1)}$ and $\vec{\Pi}^{(2)}$ into a vector $\vec{\Pi}^{(\gamma_1,\gamma_2)}$ (for the n=2 version) which is more co-linear to the conditional winning probability vector $\vec{p}^{(\gamma_1,\gamma_2)}$ than in the case of each individual game (this is just a geometric restatement of the fact that the combined game is winning). One can say that each game alternatively acts at random so as to better align these two vectors on average under the action of the other game. Choosing the previously best performing game amounts to removing this combined effect, while choosing the previously worst performing game tends to intensify this effect.

We have identified two classes of mechanisms operating in the Minority games and in the Parrondo games in which optimizing agents obtain suboptimal outcomes compared with non-optimizing agents. These examples suggest a general definition: the "illusion of control" effect occurs when low-entropy strategies (i.e. which use more information) under-perform random strategies (with maximal entropy). The illusion of control effect is related to bounded rationality as well as limited information [26] since, as we have shown, unbounded rational agents learn to converge to the symmetric mixed fully random strategies. It is only in the presence of bound rationality that agents can stick with optimization scheme on a subset of strategies. Our robust message is that, under bounded rationality, the simple (large-entropy) strategies are often to be preferred over more complex elaborated (low-entropy) strategies. This is a message that should appeal to managers and practitioners, who are well-aware in their everyday practice that simple solutions are preferable to complex ones, in the presence of the ubiquitous uncertainty.

More examples should be easy to find. For instance, control algorithms, which employ optimal parameter estimation based on past observations, have been shown to generate broad power law distributions of fluctuations and of their corresponding corrections in the control process, suggesting that, in certain situations [27], uncertainty and risk may be amplified by optimal control. In the same spirit, more quality control in code development often decreases the overall quality which itself spurs more quality control leading to a vicious circle [28]. In finance, there are many studies suggesting that most fund managers perform worse than random [29] and strong evidence that over-trading leads to anomalously large financial volatility [30]. Let us also mention the interesting experiments in which optimizing humans are found to perform worse than rats [301]. We conjecture that the illusion-of-control effect should be widespread in many strategic and optimization games and perhaps in many real life situations. Our contribution is to put this question at a quantitative level so that it can be studied rigorously to help formulate better strategies and tools for management.

**Acknowledgements**

We are grateful to Riley Crane, Gilles Daniel and especially Yannick Malevergne for helpful feedback on the manuscript.

# Appendix A. Analytic Methods and Simulations

## A1 The Minority Game: Choosing the Best Strategy

In the simplest version of the Minority Game (MG) with $N$ agents, every agent has $S = 2$ strategies and $m = 2$. In the Time Horizon Minority Game (THMG), the point (or score) table associated with strategies is not maintained from the beginning of the game and is not ever growing. It is a rolling window of finite length $\tau$ (in the simplest case $\tau = 1$). The standard MG reaches an equilibrium state after a finite number of steps $t_{st}$. At this point, the dynamics and the behavior of individual agents for a given initial quenched disorder in the MG are indistinguishable from an otherwise identical THMG with $\tau \geq t_{st}$. Extensive numerical simulations show that all the phenomenon we discuss in the THMG with the simplest parameters are found in the MG with arbitrary parameters and in the THMG with $\tau$ of generally arbitrary length and parameters. In particular, the message of our communication holds true: agents under-perform strategies.

The fundamental result of the MG is generally cast in terms of system volatility: $\sigma^2/N$. All variations of agent and strategy reward functions depend on the negative sign of the majority vote. Therefore both agent and strategy "wealth" (points, whether "real" or hypothetical) are inverse or negative functions of the volatility: The lower the value of $\sigma^2/N$, the greater the mean "wealth" of the "system", i.e., of agents. However, this mean value is scarcely ever compared to the comparable value for the raw strategies of which agents are composed. Yet agents are supposed to enhance their performance by choosing adaptively between their available strategies. In fact, the opposite is true: The optimization method is strictly a method for worsening performance.

To emphasize the relation of the MG to market-games and the illusion of optimization, we transform the fundamental result of the MG from statements on the properties of $\sigma^2/N$ to change in wealth, i.e., $\Delta W/\Delta t$ for agents and $\Delta W/\Delta t$ for strategies. We again use the simplest possible formulation—if an agent's actual (or strategy's hypothetical) vote places it in the minority, it scores $+1$ points, otherwise $-1$. Formally: At every discrete time-step $t$, each agent independently re-selects one of its $S$ strategies. It "votes" as the selected strategy dictates by taking one of two "actions," designated by a binary value:

$$a_i(t) \in \{1,0\}, \ \forall\ i,t \tag{3}$$

The state of the system as a whole at time $t$ is a mapping of the sum of all the agents' actions to the integer set $\{2N_1 - N\}$, where $N_1$ is the number of 1 votes and $N_0 = N - N_1$. This mapping is defined as:

$$A(t) = 2\sum_{i=1}^{N} a_i(t) - N = N_1 - N_0 \tag{4}$$

If $A(t) > \frac{N}{2}$, then the minority of agents will have chosen 0 at time $t$ ($N_0 < N_1$); if $A(t) < \frac{N}{2}$, then the minority of agents will have chosen 1 at time $t$ ($N_1 < N_0$). The minority choice is the "winning" decision for $t$. This is then mapped back to $\{0,1\}$:

$$D_{sys}(t) = -\text{Sgn}[A(t)] \ \therefore D_{sys}(t) \in \{-1,+1\} \rightarrow \{0,1\} \tag{5}$$

For the MG, binary strings of length $m$ form histories $\mu(t)$, with $m = \dim[\mu(t)]$. For the THMG, binary strings of length $m+\tau$ form paths (or "path histories") [8,9], with $m+\tau = \dim(\mu_t)$, where we define $\mu(t)$ as a history in the standard MG and $\mu_t$ as a path in the THMG. Then as demonstrated in [8,9], any THMG has a Markov chain formulation. For $\{m,S,N\} = \{2,2,31\}$, the typical initial quenched disorder in the strategies attributed to each of the $N$ agents is represented by the tensor $\hat{\Omega}$ and its symmetrized equivalent $\hat{\Psi} = \tfrac{1}{2}(\hat{\Omega} + \hat{\Omega}^T)$. Positions along all $S$ edges of $\hat{\Omega}$ represent an ordered listing of all available strategies. The numerical values $\Omega_{ij...}$ in $\hat{\Omega}$ indicate the number of times a specific strategy-tuple has been selected. (E.g., for two strategies per agent, $S=2$, $\Omega_{2,5}=3$ means that there are 3 agents with strategy 2 and strategy 5.) Without loss of generality, we may express $\hat{\Omega}$ in upper-triangular form since the order of strategies in a agent has no meaning. The example (6) is a typical such tensor $\hat{\Omega}$ for $S=2$, $N=31$.

$$\hat{\Omega} = \begin{pmatrix} 1 & 2 & 0 & 0 & 1 & 1 & 0 & 0 \\ 0 & 0 & 0 & 0 & 3 & 3 & 1 & 1 \\ 0 & 0 & 2 & 0 & 1 & 0 & 0 & 0 \\ 0 & 0 & 0 & 1 & 1 & 0 & 0 & 1 \\ 0 & 0 & 0 & 0 & 1 & 0 & 2 & 1 \\ 0 & 0 & 0 & 0 & 0 & 2 & 2 & 1 \\ 0 & 0 & 0 & 0 & 0 & 0 & 2 & 1 \\ 0 & 0 & 0 & 0 & 0 & 0 & 0 & 0 \end{pmatrix} \tag{6}$$

Actions are drawn from a reduced strategy space (RSS) [4,32] of dimension $2^m$. Each is associated with a strategy $k$ and a path $\mu_t$. Together they can be represented in table form as a $\dim(\text{RSS}) \times \dim(\mu_t)$ binary matrix with elements converted for convenience from $\{0,1\} \to \{-1,+1\}$, i.e., $a_k^{\mu_t} \in \{-1,+1\}$. For $m=2$, $\tau=1$, $m+\tau = \dim(\mu_t) = 3$, there are $2^3$ possible histories and $r=2^2$ reduced strategies (and $2^r$ strategies in total). In this case, the table of dimension $\dim(\text{RSS}) \times \dim(\mu_t)$ coding for all possible reduced strategies and paths reads:

$$\hat{a} \equiv \begin{pmatrix} -1 & -1 & -1 & -1 \\ -1 & -1 & +1 & +1 \\ -1 & +1 & -1 & +1 \\ -1 & +1 & +1 & -1 \\ +1 & -1 & -1 & +1 \\ +1 & -1 & +1 & -1 \\ +1 & +1 & -1 & -1 \\ +1 & +1 & +1 & +1 \end{pmatrix} \tag{7}$$

The change in wealth (point gain or loss) associated with each of the $2^r = 8$ strategies for the 8 paths (= allowed transitions between the 4 histories) at any time $t$ is then:

$$\delta \vec{S}_{\mu(t),\mu(t-1)} = \left(\hat{a}^{\mathrm{T}}\right)_{\mu(t)} \times \left\{2 Mod\left[\mu(t-1),2\right]-1\right\} \tag{8}$$

$Mod[x,y]$ is "x modulo y"; $\mu(t)$ and $\mu(t-1)$ label each of the 4 histories $\{00,01,10,11\}$ hence take on one of values $\{1,2,3,4\}$. Equation (8) picks out from (7) the correct change in wealth over a single step since the strategies are ordered in symmetrical sequence.

The change in points associated with each strategy for each of the allowed transitions between paths $\mu_t$ of the last $\tau$ time steps used to score the strategies is:

$$\vec{s}_{\mu_t} = \sum_{i=0}^{\tau-1} \delta \vec{S}_{\mu(t-i),\mu(t-i-1)} \tag{9}$$

For example, for $m = 2$ and $\tau = 1$, the strategy scores are kept for only a single time-step. There is no summation so (9) in matrix form reduces to the score:

$$\vec{s}_{\mu_t} = \delta \vec{S}_{\mu(t),\mu(t-1)} \tag{10}$$

or, listing the results for all 8 path histories:

$$\hat{\mathbf{s}}_\mu = \delta \hat{\mathbf{S}} \tag{11}$$

$\delta \hat{\mathbf{S}}$ is an 8×8 matrix that can be read as a lookup table. It denotes the change in points accumulated over $\tau = 1$ time steps for each of the 8 strategies over each of the 8 path-histories.

Instead of computing $A(t)$, we compute $A(\mu_t)$. Then for each of the $2^{m+\tau} = 8$ possible $\mu_t$, $A(\mu_t)$ is composed of a subset of wholly determined agent votes and a subset of undetermined agents whose votes must be determined by a coin toss:

$$A(\mu_t) = A_D(\mu_t) + A_U(\mu_t) \tag{12}$$

Some agents are undetermined at time t because their strategies have the same score and the tie has to be broken with a coin toss. $A_U(\mu_t)$ is a random variable characterized the binomial distribution. Its actual value varies with the number of undetermined agents. This number can be explicated *(2)*:

$$N_U(\mu_t) =$$
$$\left\{\left(1-\left[\left(\hat{a}_1^{\mathrm{T}}\right)_{(Mod[\mu_t-1,4]+1)} \otimes_\delta \left(\hat{a}_1^{\mathrm{T}}\right)_{(Mod[\mu_t-1,4]+1)}\right]\right) \circ \left(\vec{s}_{\mu_t} \otimes_\delta \vec{s}_{\mu_t}\right) \circ \hat{\Omega}\right\}_{(Mod[\mu_t-1,2^m]+1)} \tag{13}$$

"$\otimes_\delta$" is a generalized outer product, with the product being the Kronecker delta. $\vec{N}_U$ constitutes a vector of such values. The summed value of all undetermined decisions for a given $\mu_t$ is distributed binomially. Similarly *(2)*:

$$A_D(\mu_t) =$$
$$\left(\sum_{r=1}^{8} \left\{\left[\left(1 - Sgn\left[\vec{s}_{\mu_t} \ominus \vec{s}_{\mu_t}\right]\right) \circ \hat{\Psi}\right] \bullet \hat{a}_1\right\}_r\right)_{(Mod[\mu_t-1,2^m]+1)} \tag{14}$$

An example of how (13) and (14) can be deduced is given later in the context of the original definition of alternate types of agents. Details may also be found in Ref.[9]. We define $\vec{A}_D$ as a vector of the determined contributions to $A(t)$ for each path $\mu_t$. In expression (14) $\mu_t$ numbers paths from 1 to 8 and is therefore here an index. $\vec{s}_{\mu_t}$ is the "$\mu_t$<sup>th</sup>" vector of net point gains or losses for each strategy when at $t$ the system has traversed the path $\mu_t$ (i.e., it is the "$\mu_t$<sup>th</sup>" element of the matrix $\hat{\mathbf{s}}_\mu = \delta\hat{\mathbf{S}}$ in (11)). "⊖" is a generalized outer product of two vectors with subtraction as the product. The two vectors in this instance are the same, i.e., $\vec{s}_{\mu_t}$. "∘" is Hadamard (element-by-element) multiplication and "•" the standard inner product. The index $r$ refers to strategies in the RSS. Summation over $r$ transforms the base-ten code for $\mu_t$ into $\{1,2,3,4,1,2,3,4\}$. Selection of the proper number is indicated by the subscript expression on the entire right-hand side of (13). This expression yields an index number, i.e., selection takes place 1 + Modulo 4 with respect to the value of $(\mu_t - 1)$.

To obtain the transition matrix for the system as a whole, we require the $2^{m+\tau} \times 2^{m+\tau}$ adjacency matrix that filters out disallowed transitions. Its elements are $\Gamma_{\mu_t, \mu_{t-1}}$:

$$\hat{\mathbf{\Gamma}} = \begin{pmatrix} 1 & 0 & 0 & 0 & 1 & 0 & 0 & 0 \\ 1 & 0 & 0 & 0 & 1 & 0 & 0 & 0 \\ 0 & 1 & 0 & 0 & 0 & 1 & 0 & 0 \\ 0 & 1 & 0 & 0 & 0 & 1 & 0 & 0 \\ 0 & 0 & 1 & 0 & 0 & 0 & 1 & 0 \\ 0 & 0 & 1 & 0 & 0 & 0 & 1 & 0 \\ 0 & 0 & 0 & 1 & 0 & 0 & 0 & 1 \\ 0 & 0 & 0 & 1 & 0 & 0 & 0 & 1 \end{pmatrix} \quad (15)$$

Equations (13), (14) and (15) yield the history-dependent $(m+\tau) \times (m+\tau)$ matrix $\hat{\mathbf{T}}$ with elements $T_{\mu_t, \mu_{t-1}}$, representing the 16 allowed probabilities of transitions between the two sets of 8 path-histories $\mu_t$ and $\mu_{t-1}$:

$$T_{\mu_t, \mu_{t-1}} = \Gamma_{\mu_t, \mu_{t-1}} \times$$
$$\sum_{x=0}^{N_U(\mu_t)} \left\{ \binom{N_U(\mu_t)}{x} \left(\frac{1}{2}\right)^{N_U(\mu_t)} \times \delta\left[\text{Sgn}\left(A_D(\mu_t) + 2x - N_U(\mu_t)\right) + \left(2\,\text{Mod}\{\mu_{t-1}, 2\} - 1\right)\right] \right\} \quad (16)$$

The expression $\binom{N_U(\mu_t)}{x}\left(\frac{1}{2}\right)^{N_U(\mu_t)}$ in (16) represents the binomial distribution of undetermined outcomes under a fair coin-toss with mean = $A_D(\mu_t)$. Given a specific $\hat{\mathbf{\Omega}}$,

$$\langle A(\mu_t) \rangle = A_D(\mu_t) \; \forall \; \mu_t \quad (17)$$

We now tabulate the number of times each strategy is represented in $\hat{\mathbf{\Omega}}$, regardless of coupling (i.e., of which strategies are associated in forming agent S-tuples):

$$\vec{\kappa} \equiv \sum_{k=1}^{2^{m+\tau}} \left(\hat{\mathbf{\Omega}} + \mathbf{\Omega}^\top\right)_k = 2\sum_{k=1}^{2^{m+\tau}} \hat{\mathbf{\Psi}}_k = \left\{n(\sigma_1), n(\sigma_2), \ldots n(\sigma_{2^{m+\tau}})\right\} \quad (18)$$

where $\sigma_k$ is the $k^{th}$ strategy in the RSS, $\hat{\Omega}_k, \hat{\Omega}_k^\top$ and $\hat{\Psi}_k$ are the $k^{th}$ element (vector) in each tensor and $n(\sigma_k)$ represents the number of times $\sigma_k$ is present across all strategy tuples. Therefore

$$\langle \Delta W_{Agent} \rangle = -\tfrac{1}{N} Abs(\vec{A}_D) \cdot \vec{\mu} \tag{19}$$

and

$$\langle \Delta W_{Strategy} \rangle = \tfrac{1}{2N} (\hat{\mathbf{s}}_\mu \cdot \vec{\kappa}) \cdot \vec{\mu} \tag{20}$$

with $\vec{\mu}$ the normalized steady-state probability vector for $\hat{\mathbf{T}}$. Expression (19) states that the mean per-step change in wealth for agents equals $-1$ times the probability-weighted sum of the (absolute value of the) *determined* vote imbalance associated with a given history. Expression (20) states that the mean per-step change in wealth for individual strategies equals the probability-weighted sum of the representation of each strategy (in a given $\hat{\Omega}$) times the sum over the per-step wealth change associated with every history. The $-1$ in (19) reflects the minority rule. I.e., the awarding of points is the negative of the direction of the vote imbalance. No minus sign is required in (20) as it is already accounted for in (7).

**Figure A1** shows the cumulative mean change in wealth for strategies versus agents over time, given (15).

As first studied in [11,12], and discussed in the body of the manuscript, agent performance is inversely proportional to the Hamming distance between strategies within agents. With the variation expected of a single example, our sample $\hat{\Omega}$ given by (6) reproduces this relation as shown in **Figure A2**. Thus agent performance is distributed within $\hat{\Omega}$ in orderly if complex fashion. The mean over many $\hat{\Omega}$ corresponds to a "flat" $\hat{\Omega}$.

### A2 The Minority Game: Choosing the Worst Strategy

First, we re-cast the initial quenched disorder on the set of strategies attributed to the $N$ agents in a given game realization as a two-component tensor: $\hat{\Omega} = \{\hat{\Omega}^+, \hat{\Omega}^-\}$. $\hat{\Omega}^+$ represents standard (S) agents that adapt as before; $\hat{\Omega}^-$ represents "counteradaptive" (C) agents that instead select their worst-performing strategies. In our example (6) then, suppose we select at random 3 agents to use the C rule, one each at $\Omega_{1,2}, \Omega_{2,6}$ and $\Omega_{7,8}$:

$$\hat{\Omega} = \{\hat{\Omega}^+, \hat{\Omega}^-\} =$$

$$\left\{ \begin{pmatrix} 1 & 1 & 0 & 0 & 1 & 1 & 0 & 0 \\ 0 & 0 & 0 & 0 & 3 & 2 & 1 & 1 \\ 0 & 0 & 2 & 0 & 1 & 0 & 0 & 0 \\ 0 & 0 & 0 & 1 & 1 & 0 & 0 & 1 \\ 0 & 0 & 0 & 0 & 1 & 0 & 2 & 1 \\ 0 & 0 & 0 & 0 & 0 & 2 & 2 & 1 \\ 0 & 0 & 0 & 0 & 0 & 0 & 2 & 0 \\ 0 & 0 & 0 & 0 & 0 & 0 & 0 & 0 \end{pmatrix}, \begin{pmatrix} 0 & 1 & 0 & 0 & 0 & 0 & 0 & 0 \\ 0 & 0 & 0 & 0 & 0 & 1 & 0 & 0 \\ 0 & 0 & 0 & 0 & 0 & 0 & 0 & 0 \\ 0 & 0 & 0 & 0 & 0 & 0 & 0 & 0 \\ 0 & 0 & 0 & 0 & 0 & 0 & 0 & 0 \\ 0 & 0 & 0 & 0 & 0 & 0 & 0 & 0 \\ 0 & 0 & 0 & 0 & 0 & 0 & 0 & 1 \\ 0 & 0 & 0 & 0 & 0 & 0 & 0 & 0 \end{pmatrix} \right\} \quad (21)$$

For any number of C agents in $\hat{\Omega}$ thus redefined, the analytic expression for $\hat{T}$ need only be modified by decomposing $A_D(\mu_t)$ accordingly. The new term in $A_D(\mu_t)$ makes evident the symmetry of the C rule with respect to the S rule, and the lack of privilege of C agents. Thus:

$$A_D(\mu_t) = \left( \sum_{r=1}^{8} \left\{ \left[ \left(1 - Sgn[\bar{s}_{\mu_t} \ominus \bar{s}_{\mu_t}]\right) \circ \hat{\Psi}^+ + \left(1 + Sgn[\bar{s}_{\mu_t} \ominus \bar{s}_{\mu_t}]\right) \circ \hat{\Psi}^- \right] \bullet \hat{a}_1 \right\}_r \right)_{(\text{Mod}[\mu_t - 1, 2^m] + 1)} \quad (22)$$

with

$$\hat{\Psi}^+ = \tfrac{1}{2}\left(\hat{\Omega}^+ + \hat{\Omega}^{+\mathrm{T}}\right); \quad \hat{\Psi}^- = \tfrac{1}{2}\left(\hat{\Omega}^- + \hat{\Omega}^{-\mathrm{T}}\right) \quad (23)$$

The number of undetermined agent votes remains unchanged. In (13), $\hat{\Omega}$ need only be replaced with $\left(\hat{\Omega}^+ + \hat{\Omega}^-\right)$:

$$N_U(\mu_t) = \left\{ \left(1 - \left[\left(\hat{a}_1^\mathrm{T}\right)_{(\text{Mod}[\mu_t - 1, 4] + 1)} \otimes_\delta \left(\hat{a}_1^\mathrm{T}\right)_{(\text{Mod}[\mu_t - 1, 4] + 1)} \right] \right) \circ \left(\bar{s}_{\mu_t} \otimes_\delta \bar{s}_{\mu_t}\right) \circ \left(\hat{\Omega}^+ + \hat{\Omega}^-\right) \right\}_{(\text{Mod}[\mu_t - 1, 2^m] + 1)} \quad (24)$$

Results for numerical simulation and analytic calculation are in close agreement even for a single short simulation, as illustrated in

**Table A1**.

The 3 C agents of 31 now perform so well that they significantly raise the overall performance of the system as detailed in **Figure A3**. They not only outperform both their own strategies and the other S agents on average, they generate net positive gain. The *hypothetical* outperformance of unused relative to used strategies in the MG was first observed in *(5)*. But the explicit generation of positive results, by agents simply deploying their unused strategies (without privileging), has not been tested. (In the case of $S = 2$, "unused" are by definition the "worst-performing".)

We discuss in the manuscript and elsewhere the phenomena that arise as the proportion of S and C agents are varied for different parameters of the MG. We emphasize here only the fact that extensive numerical studies confirm that the phenomenon here illustrated persist over a very wide range of parameters for both the MG and the THMG.

## A3 The Minority Game: Random Agents

We provide in **Figure A4** some numerical results for a MG game with N agents total, $N_R$ of which employ such a fully random symmetric choice. The remaining $N_S = N-N_R$ "special" agents (with $N_R \gg N_S$) will all be one of two possible types: (i) agents with S fixed strategies that choose their worst performing strategy to make the decision at the next step (referred to above as anti-optimizing); (ii) agents with a single fixed strategy. We use the simplest example, that of $N_S = 1$ (with $\tau = 1$), to illustrate the fact that in the MG, agents allowed/restricted to a fully symmetric random choice outperform agents that attempt to optimize. (Note that the outperformance and absolute positive returns associated with a small proportion of anti-optimizing agents, requires *the remaining agents to optimize*, as described above. Here the small proportion of or optimizing and anti-optimizing agents compete with fully random agents.)

**Figures and Tables**

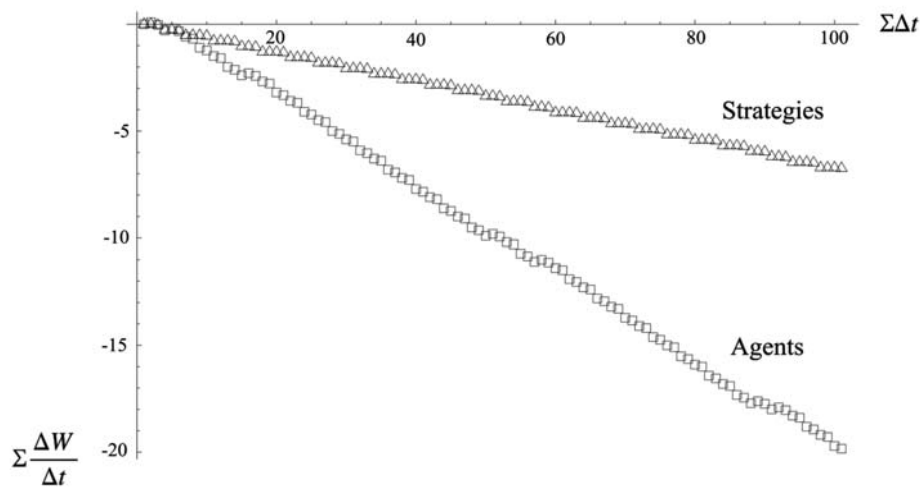

**Figure A1**: Mean Strategy versus Agent Cumulative Change in Wealth in the THMG. $\{m, S, N\} = \{2, 2, 31\}$; 100 time steps

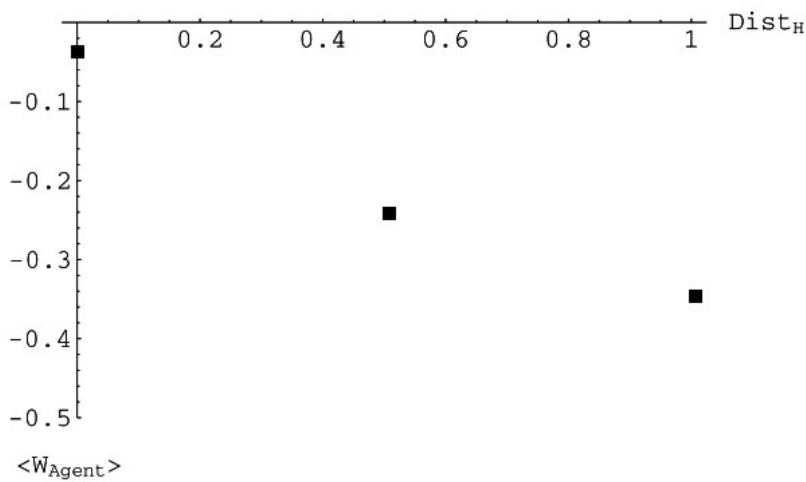

**Figure A2**: Agent wealth as a function of Hamming distance between strategy pairs in agents for the example simulation.

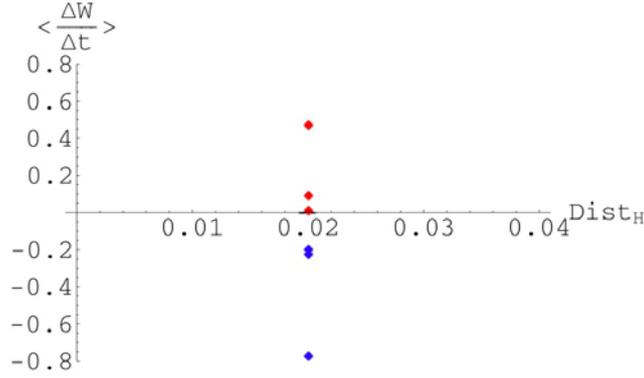

**Figure A3**: Average wealth variation per time step for different agents. In red are shown the wealth variations of the three among the 31 agents which use counteradaptive ("C", choose worst) strategy selection. The usual underperformance of agents compared to individual strategies when using standard selection rule ("S", choose best) is shown in the blue dots.

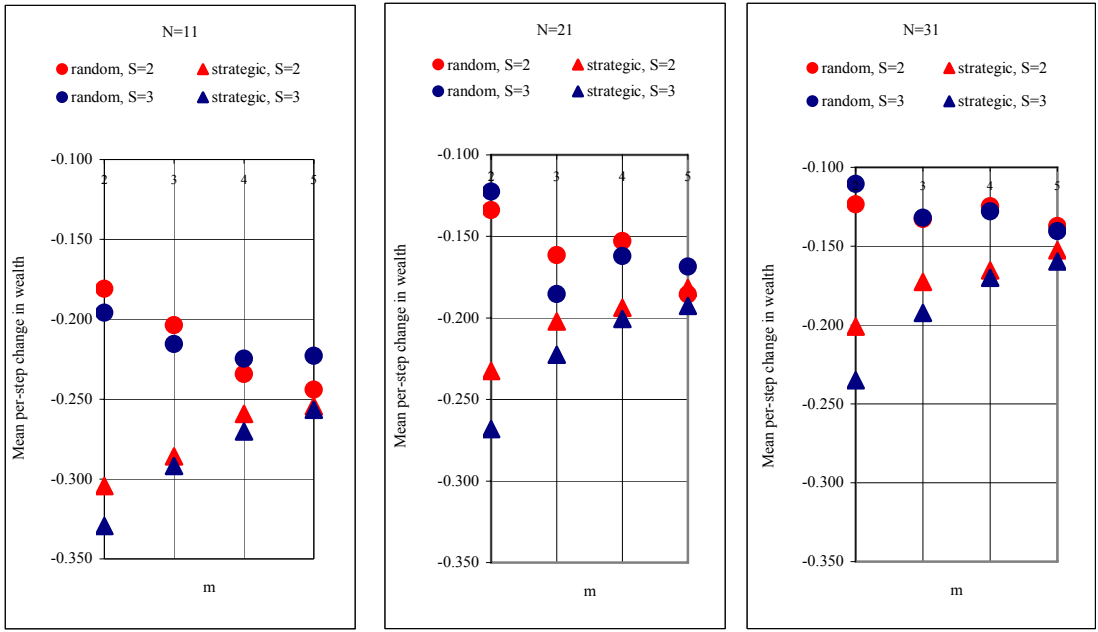

**Figure A4**: Performance (mean change in wealth per step) of a single optimizing agent versus all other agents making a symmetric random choice in a MG-like game. From left to right n=11, 21, 31. S=2,3 m= 2,3,4,5 and τ=1. Random agents always outperform optimizing agents. Similar results obtain for other values of n, m, S and τ. Within statistical fluctuations typical for the number of runs/random selection of strategies comprising the optimizing agent (100 runs), results for anti-optimizing agents are identical.

**Table A1**: Numerical/Analytic Results of THMG with and without 3 C Agents 28 S Agents

|         | $\langle \Delta W_{Agent} \rangle$ | $\langle \Delta W_{Strategy} \rangle$ |
|---------|------------------------------------|---------------------------------------|
| *With*    | –0.14/–0.14                        | –0.05/–0.05                           |
| *Without* | –0.26/–0.26                        | –0.05/–0.05                           |